\documentclass[aps, prl, superscriptaddress, nofootinbib, 
               amsmath, amssymb, twocolumn,
               preprintnumbers, showpacs,
               raggedbottom,
               floatfix,
               ]{revtex4-1}
\usepackage[utf8]{inputenc}
\usepackage[aps]{my_paper}                       
\usepackage{my_acronyms}                         

\newcommand{\h}{l_{\text{coh}}}

\newcommand{\E}{\mathrm{e}}
\renewcommand{\E}{e}

\begin{document}
 
\title{Vortex Rings in a Trap}

\author{Aurel Bulgac}
\email{bulgac@uw.edu}
\affiliation{Department of Physics, University of Washington, Seattle,
  Washington 98195--1560, USA}

\author{Michael McNeil Forbes}
\email{mforbes@alum.mit.edu}
\affiliation{Institute for Nuclear Theory, University of Washington,
  Seattle, Washington 98195--1550, USA}
\affiliation{Department of Physics, University of Washington, Seattle,
  Washington 98195--1560, USA}
\affiliation{Department of Physics \& Astronomy, Washington State University,
  Pullman, Washington 99164--2814, USA}

\begin{abstract}
\noindent
We present a simple Hamiltonian description of the dynamics of a quantized
vortex ring in a trapped superfluid, compare this description with dynamical
simulations, and characterize the dependence of the dynamics of the shape of the
trap.
\end{abstract}

\date{\today}

\pacs{03.75.Lm, 
03.75.Kk, 
67.85.De, 
67.85.-d,  
}

\preprint{INT-PUB-13-048}
\preprint{NT@UW-13-29}

\maketitle

\noindent
Topological excitations in superfluid systems have fascinated researchers for
decades.  Quantized vortices and quantized vortex rings in particular exhibit
unusual and complicated dynamics that generally require complicated simulations.
Sometimes one can treat such objects in a simplified manner as a particle with
an unusual dispersion relation, allowing for a more intuitive understanding of
the dynamics.  Here we extend on a recent preprint by
Pitaevskii~\cite{Pitaevskii:2013} that attempts to provide a simple theoretical
formalism describing the motion of a quantized vortex ring in a trapped
superfluids.  We correct several deficiencies in that description, and show how
the corrected hydrodynamic theory explains aspects of the \gls{MIT}
experiment~\cite{Yefsah:2013} that observes long-lived and slowly moving ``heavy
solitons.''  This complements the numerical explanation~\cite{Bulgac:2013d}
which suggested that the experiment sees vortex rings: an idea supported
by~\cite{Cetoli:2013, Reichl:2013, Wen:2013}.

We start with the kinetic energy $E_R$, the linear momentum $P$, and the
velocity $v$ of a single vortex ring of radius $R$ in an infinite medium
(see~\cite{Roberts:2001} for a review):
\begin{subequations}
  \begin{align}
    E_R &= \frac{1}{2} m n R\kappa^2\ln\frac{8R}{\E a}, \label{energy} \\
    P &= mn \kappa \pi R^2, \label{momentum} \\
    v &= \pdiff{E_R}{P}=\frac{\kappa}{4\pi R}\ln\frac{8R}{a}. \label{velocity}
  \end{align}
\end{subequations}
The vortex has a core size $a\sim \h$ which parameterizes the microscopic
structure of the vortex, and winding $\kappa=\pi\hbar/m$.  The spirit of
Pitaevskii's model~\cite{Pitaevskii:2013} is that for thin vortices in a large
trap, the same relationships can be used where the number density $n(R,Z)$ now
depends on the position $Z$ of the vortex ring.  What is missing from his
formulation, however, is the potential energy due to the trapping potential
$V_{\text{ext}}(\vect{r})$.

Within this framework, the correct energy dependence must be described by the
sum of both kinetic and potential terms $E = E_R + V$:
\begin{multline}
  E(R,Z) = \frac{ \pi^2 \hbar^2 }{2m}n(R,Z)R\ln\frac{8R}{\E a} +\\
  - \beta \, 2\pi^2 R \,a^2 \, n(R,Z)\,V_{\text{ext}}(R,Z).
  \label{eq:energy}
\end{multline}
The second term, missing in Ref.~\cite{Pitaevskii:2013}, is crucial for a
correct description of the physics.  It describes buoyancy: the change in energy
of the system due to the position of the density depletion in the vortex core.
We estimate the depletion as some fraction $\beta \sim 1$ of the volume of an
annular tube of radius $R$ and cross-sectional area $\pi a^2$, which must also
be fit or calculated from a microscopic description.
Pitaevskii~\cite{Pitaevskii:2013} determines the radius and the position of the
vortex ring from the energy conservation $E(R,Z)=\text{const}$, and the dynamics
from the velocity~\eqref{velocity}.

The importance of the missing potential term is clearly demonstrated by
considering an incompressible superfluid liquid for which $n(R,Z)=\text{const}$.
In this case, the formalism would assert that a vortex ring moves with constant
velocity and constant radius, irrespective of the form of the external
potential.  It is the missing potential term that generates interesting
dynamics: the depletion will give rise to a buoyant force (as with an air bubble
in water) that induces perpendicular motion via the well-known Magnus
relationship.

Another contribution neglected in~\cite{Pitaevskii:2013} is due to boundary
effects. In close proximity to a sharp surface -- either a hard wall or the
surface of a liquid -- the velocity \eqref{velocity} will be altered in a way
that can be calculated by including appropriately located ``image'' vortices
(see for example~\cite{Schwarz:1985}) that characterize the vanishing velocity
at the boundary.  In two-dimensions, for example, a vortex close to a hard wall
will move along the wall as if in the presence of an anti-vortex located on the
other side of the wall.  The effects of sharp boundaries and surfaces may thus
be included in the Hamiltonian formulation by adding these images to the
energy~\eqref{eq:energy}, though care must be taken in locating these images to
account for finite size effects as discussed in~\cite{Mason:2006}.  The
Hamiltonian description~\eqref{eq:energy} thus augmented with images acts as a
``vortex filament model''~\cite{Schwarz:1985} for vortex rings in an
compressible fluid.

We proceed by considering the \gls{UFG} whose equation of state can be expressed
in terms of the dimensionless Bertsch parameter $\xi$ through the energy density
$\mathcal{E} = \tfrac{3}{5}\xi n\varepsilon_{F}$ with the Fermi energy
$\varepsilon_F = \hbar^2k_F^2/2m$ expressed in terms of the Fermi wavevector
$k_F = (3\pi^2 n)^{1/3}$.  Since there are no length-scales for the \gls{UFG},
the vortex core-size $a = \alpha/k_F$ where $\alpha$ is of order unity. The
contours of constant energy describing the motion of a trapped vortex ring are
thus described by contours of
\begin{gather}
  E(R,Z) = \frac{\pi^2\hbar^2 }{2m}R n \left[
    \ln\frac{8k_FR}{\E\alpha}
    - 2\xi\alpha^2\beta\frac{V_{\text{ext}}}{\mu -V_{\text{ext}}}
  \right]
\end{gather}
where $n(R,Z)$, $k_F(R,Z)$, and $V_{\text{ext}}(R,Z)$ 
are local quantities and vary in space according to the
\gls{TF} approximation $\xi \varepsilon_F(R,Z) + V_{\text{ext}}(R,Z) = \mu$
\begin{gather}
  n(R,Z) =
  \frac{1}{3\pi^2} \left[
    \frac{2m (\mu -V_{\text{ext}}(R,Z))}{\hbar^2\xi}
  \right]^{3/2}.
\end{gather}
The parameters $\alpha$ and $\beta$ must be adjusted to describe the size
and filling of the vortex core which must be determined from measurement or
microscopic calculations: for the \gls{UFG} they should be close to unity.

\begin{figure}[tbp]
  \includegraphics[width=\columnwidth]{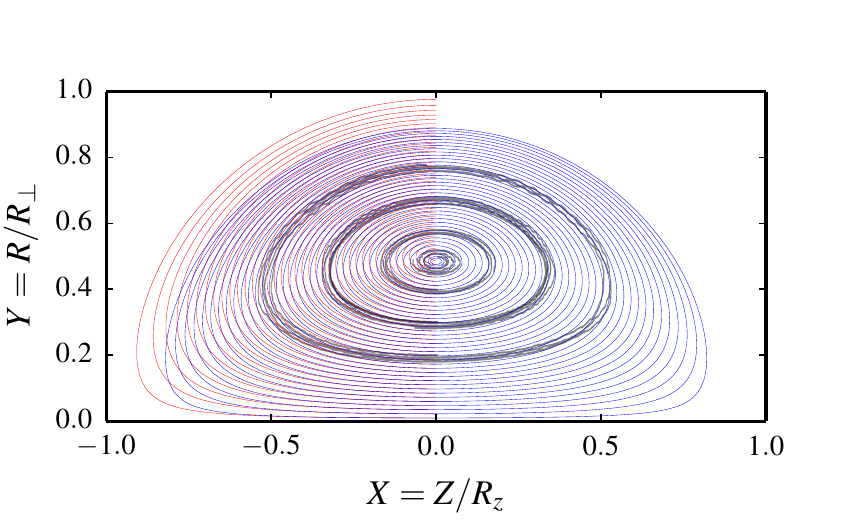}
  \caption{\label{fig:contours}
    Constant-energy contours (thin blue curves) compared trajectories from the
    simulations in~\cite{Bulgac:2013d} (thick black curves: note small
    oscillations are due to the presence of phonons).  To centre the
    trajectories for comparison with the data we find that $2\xi \alpha^2\beta
    \approx 1.3$ (we have kept the full dependence of $L(X,Y)$ while generating
    these contours).  The trajectories from~\cite{Pitaevskii:2013} are included
    on the left (red), clearly showing the effect of the missing potential
    (buoyancy).}
\end{figure}

We now consider vortex rings in a harmonic trap as were show
in~\cite{Bulgac:2013d} as able to describe the \gls{MIT}
experiment~\cite{Yefsah:2013}:
\begin{gather}
  V_{\text{ext}} (R,Z)= \mu\left(
    \frac{R^2}{R_\perp^2} + \frac{Z^2}{R_z^2}\right) = \mu(Y^2 + X^2)
\end{gather}
where we have used the notations of~\cite{Pitaevskii:2013} with dimensionless
coordinates $Y = R/R_\perp$ and $X=Z/R_z$ expressed in terms of the \gls{TF}
radii $R_{\perp,z} = \sqrt{2\mu/m}/\omega_{\perp,z}$ so that $n(X,Y) = n_0(1-X^2
- Y^2)^{3/2}$.  The energy of a vortex ring may thus be expressed as $E(R,Z) =
(\pi^2\hbar^2 R_\perp n_0 L/2m)f(Y,X)$ where $L=\ln (8R/\E a)$ is approximately
constant, but needs to be treated explicitly for quantitative results:
\begin{multline}
  f(Y,X) = Y(1-X^2-Y^2)^{\frac{3}{2}} \\
  \times \left[
    1
    - \frac{2\xi\alpha^2\beta}{L}\frac{X^2+Y^2}{1-X^2-Y^2} 
  \right].
  \label{eq:13}
\end{multline}
The resulting trajectories are shown in Fig.~\ref{fig:contours} where they are
compared with both the results of Pitaevskii~\cite{Pitaevskii:2013} and the full
dynamical simulations of~\cite{Bulgac:2013d}.

The second term of~\eqref{eq:13} arising from the potential is missing from
Eq.~(13) of Ref.~\cite{Pitaevskii:2013}, but is somewhat small for the
harmonically trapped \gls{UFG}: for example, in the \gls{MIT}
experiment~\cite{Yefsah:2013}, one has $k_FR \approx 20$, hence the coefficient
is $L\approx 5$, while $2\xi \alpha^2\beta =\mathcal{O}(1)$.  This explains the
qualitative behaviour of the results in~\cite{Pitaevskii:2013}.  On the other
hand, for an incompressible liquid one must replace $n_0(1-X^2-Y^2)^{3/2}$ with
a constant $n_0$ and~\eqref{eq:13} becomes $f(Y,X)= Y-2\xi\alpha^2\beta
Y(X^2+Y^2)/L$ where $Y^2$ (up to a multiplicative constant) is the momentum
canonically conjugate to coordinate $X$: without the potential, the dynamics
would completely incorrect.

Using Eq.~\eqref{momentum} and introducing the dimensionless time $\tau = t\,
2m/ \hbar R_z R_\perp$, one can write the action for a vortex ring in trap
(using non-canonical coordinates $X$ and $Y$)
\begin{multline}
  S = \hbar \pi^2 n_0R_\perp^2R_z\times\\
  \times \int \left[ Y^2(1-X^2-Y^2)^{3/2}dX
  -Lf(Y,X)d\tau\right].
\end{multline}
With the exception of a weak dependence through the logarithmic term $L$, it
follows that the $X(\tau),\,Y(\tau)$ coordinates the trajectories are
independent on the shape and size of the cloud.  Hence, the real period of
oscillations is proportional to $\lambda = R_z/R_\perp$, consistent with the
numerical results of Ref.~\cite{Bulgac:2013d} and seen in the
experiment~\cite{Yefsah:2013}. The dependence of the oscillation period on the
vortex radius ($T\propto R$ up to logarithmic corrections, for fixed $R_z$) is
thus in agreement with the estimate derived in Ref.~\cite{Bulgac:2013d} and
disagreement with Ref.~\cite{Pitaevskii:2013}, where it was determined that
$T\propto R^{2/3}$.

We acknowledge support under U.S. Department of Energy (DoE) Grants
Nos\@. DE-FG02-97ER41014 and DE-FG02-00ER41132.


\begin{thebibliography}{9}%
\makeatletter
\providecommand \@ifxundefined [1]{%
 \@ifx{#1\undefined}
}%
\providecommand \@ifnum [1]{%
 \ifnum #1\expandafter \@firstoftwo
 \else \expandafter \@secondoftwo
 \fi
}%
\providecommand \@ifx [1]{%
 \ifx #1\expandafter \@firstoftwo
 \else \expandafter \@secondoftwo
 \fi
}%
\providecommand \natexlab [1]{#1}%
\providecommand \enquote  [1]{``#1''}%
\providecommand \bibnamefont  [1]{#1}%
\providecommand \bibfnamefont [1]{#1}%
\providecommand \citenamefont [1]{#1}%
\providecommand \href@noop [0]{\@secondoftwo}%
\providecommand \href [0]{\begingroup \@sanitize@url \@href}%
\providecommand \@href[1]{\@@startlink{#1}\@@href}%
\providecommand \@@href[1]{\endgroup#1\@@endlink}%
\providecommand \@sanitize@url [0]{\catcode `\\12\catcode `\$12\catcode
  `\&12\catcode `\#12\catcode `\^12\catcode `\_12\catcode `\%12\relax}%
\providecommand \@@startlink[1]{}%
\providecommand \@@endlink[0]{}%
\providecommand \url  [0]{\begingroup\@sanitize@url \@url }%
\providecommand \@url [1]{\endgroup\@href {#1}{\urlprefix }}%
\providecommand \urlprefix  [0]{URL }%
\providecommand \Eprint [0]{\href }%
\providecommand \doibase [0]{http://dx.doi.org/}%
\providecommand \selectlanguage [0]{\@gobble}%
\providecommand \bibinfo  [0]{\@secondoftwo}%
\providecommand \bibfield  [0]{\@secondoftwo}%
\providecommand \translation [1]{[#1]}%
\providecommand \BibitemOpen [0]{}%
\providecommand \bibitemStop [0]{}%
\providecommand \bibitemNoStop [0]{.\EOS\space}%
\providecommand \EOS [0]{\spacefactor3000\relax}%
\providecommand \BibitemShut  [1]{\csname bibitem#1\endcsname}%
\let\auto@bib@innerbib\@empty
\bibitem [{\citenamefont {Pitaevskii}(2013)}]{Pitaevskii:2013}%
  \BibitemOpen
  \bibfield  {author} {\bibinfo {author} {\bibfnamefont {L.~P.}\ \bibnamefont
  {Pitaevskii}},\ }\href {http://arxiv.org/abs/1311.4693} {\enquote {\bibinfo
  {title} {{H}ydrodynamic theory of motion of quantized vortex rings in trapped
  superfluid gases},}\ } (\bibinfo {year} {2013}),\ \Eprint
  {http://arxiv.org/abs/arXiv:1311.4693} {arXiv:1311.4693} \BibitemShut
  {NoStop}%
\bibitem [{\citenamefont {Yefsah}\ \emph {et~al.}(2013)\citenamefont {Yefsah},
  \citenamefont {Sommer}, \citenamefont {Ku}, \citenamefont {Cheuk},
  \citenamefont {Ji}, \citenamefont {Bakr},\ and\ \citenamefont
  {Zwierlein}}]{Yefsah:2013}%
  \BibitemOpen
  \bibfield  {author} {\bibinfo {author} {\bibfnamefont {T.}~\bibnamefont
  {Yefsah}}, \bibinfo {author} {\bibfnamefont {A.~T.}\ \bibnamefont {Sommer}},
  \bibinfo {author} {\bibfnamefont {M.~J.~H.}\ \bibnamefont {Ku}}, \bibinfo
  {author} {\bibfnamefont {L.~W.}\ \bibnamefont {Cheuk}}, \bibinfo {author}
  {\bibfnamefont {W.}~\bibnamefont {Ji}}, \bibinfo {author} {\bibfnamefont
  {W.~S.}\ \bibnamefont {Bakr}}, \ and\ \bibinfo {author} {\bibfnamefont
  {M.~W.}\ \bibnamefont {Zwierlein}},\ }\href {\doibase 10.1038/nature12338}
  {\bibfield  {journal} {\bibinfo  {journal} {Nature}\ }\textbf {\bibinfo
  {volume} {499}},\ \bibinfo {pages} {426} (\bibinfo {year} {2013})},\ \Eprint
  {http://arxiv.org/abs/1302.4736} {arXiv:1302.4736 [cond-mat]} \BibitemShut
  {NoStop}%
\bibitem [{\citenamefont {Bulgac}\ \emph {et~al.}(2013)\citenamefont {Bulgac},
  \citenamefont {Forbes}, \citenamefont {Kelley}, \citenamefont {Roche},\ and\
  \citenamefont {Wlaz{\l}owski}}]{Bulgac:2013d}%
  \BibitemOpen
  \bibfield  {author} {\bibinfo {author} {\bibfnamefont {A.}~\bibnamefont
  {Bulgac}}, \bibinfo {author} {\bibfnamefont {M.~M.}\ \bibnamefont {Forbes}},
  \bibinfo {author} {\bibfnamefont {M.~M.}\ \bibnamefont {Kelley}}, \bibinfo
  {author} {\bibfnamefont {K.~J.}\ \bibnamefont {Roche}}, \ and\ \bibinfo
  {author} {\bibfnamefont {G.}~\bibnamefont {Wlaz{\l}owski}},\ }\href
  {http://arxiv.org/abs/1306.4266} {\enquote {\bibinfo {title} {Quantized
  superfluid vortex rings in the unitary fermi gas},}\ } (\bibinfo {year}
  {2013}),\ \Eprint {http://arxiv.org/abs/1306.4266} {arXiv:1306.4266
  [cond-mat.quant-gas]} \BibitemShut {NoStop}%
\bibitem [{\citenamefont {Cetoli}\ \emph {et~al.}(2013)\citenamefont {Cetoli},
  \citenamefont {Brand}, \citenamefont {Scott}, \citenamefont {Dalfovo},\ and\
  \citenamefont {Pitaevskii}}]{Cetoli:2013}%
  \BibitemOpen
  \bibfield  {author} {\bibinfo {author} {\bibfnamefont {A.}~\bibnamefont
  {Cetoli}}, \bibinfo {author} {\bibfnamefont {J.}~\bibnamefont {Brand}},
  \bibinfo {author} {\bibfnamefont {R.~G.}\ \bibnamefont {Scott}}, \bibinfo
  {author} {\bibfnamefont {F.}~\bibnamefont {Dalfovo}}, \ and\ \bibinfo
  {author} {\bibfnamefont {L.~P.}\ \bibnamefont {Pitaevskii}},\ }\href
  {\doibase 10.1103/PhysRevA.88.043639} {\bibfield  {journal} {\bibinfo
  {journal} {Phys. Rev. A}\ }\textbf {\bibinfo {volume} {88}},\ \bibinfo
  {pages} {043639} (\bibinfo {year} {2013})},\ \Eprint
  {http://arxiv.org/abs/1307.3717} {arXiv:1307.3717} \BibitemShut {NoStop}%
\bibitem [{\citenamefont {Reichl}\ and\ \citenamefont
  {Mueller}(2013)}]{Reichl:2013}%
  \BibitemOpen
  \bibfield  {author} {\bibinfo {author} {\bibfnamefont {M.~D.}\ \bibnamefont
  {Reichl}}\ and\ \bibinfo {author} {\bibfnamefont {E.~J.}\ \bibnamefont
  {Mueller}},\ }\href {\doibase 10.1103/PhysRevA.88.053626} {\bibfield
  {journal} {\bibinfo  {journal} {Phys. Rev. A}\ }\textbf {\bibinfo {volume}
  {88}},\ \bibinfo {pages} {053626} (\bibinfo {year} {2013})},\ \Eprint
  {http://arxiv.org/abs/1309.7012} {arXiv:1309.7012} \BibitemShut {NoStop}%
\bibitem [{\citenamefont {Wen}\ \emph {et~al.}(2013)\citenamefont {Wen},
  \citenamefont {Zhao},\ and\ \citenamefont {Ma}}]{Wen:2013}%
  \BibitemOpen
  \bibfield  {author} {\bibinfo {author} {\bibfnamefont {W.}~\bibnamefont
  {Wen}}, \bibinfo {author} {\bibfnamefont {C.}~\bibnamefont {Zhao}}, \ and\
  \bibinfo {author} {\bibfnamefont {X.}~\bibnamefont {Ma}},\ }\href
  {http://arxiv.org/abs/1309.7408} {\enquote {\bibinfo {title} {{D}ark solitons
  dynamics and snake instability in superfluid {F}ermi gases trapped by an
  anisotropic harmonic potential},}\ } (\bibinfo {year} {2013}),\ \Eprint
  {http://arxiv.org/abs/arXiv:1309.7408} {arXiv:1309.7408} \BibitemShut
  {NoStop}%
\bibitem [{\citenamefont {Roberts}\ and\ \citenamefont
  {Berloff}(2001)}]{Roberts:2001}%
  \BibitemOpen
  \bibfield  {author} {\bibinfo {author} {\bibfnamefont {P.}~\bibnamefont
  {Roberts}}\ and\ \bibinfo {author} {\bibfnamefont {N.}~\bibnamefont
  {Berloff}},\ }in\ \href {\doibase 10.1007/3-540-45542-6_23} {\emph {\bibinfo
  {booktitle} {Quantized Vortex Dynamics and Superfluid Turbulence}}},\
  \bibinfo {series} {Lecture Notes in Physics}, Vol.\ \bibinfo {volume} {571},\
  \bibinfo {editor} {edited by\ \bibinfo {editor} {\bibfnamefont {C.~F.}\
  \bibnamefont {Barenghi}}, \bibinfo {editor} {\bibfnamefont {R.~J.}\
  \bibnamefont {Donnelly}}, \ and\ \bibinfo {editor} {\bibfnamefont {W.~F.}\
  \bibnamefont {Vinen}}}\ (\bibinfo  {publisher} {Springer-Verlag},\ \bibinfo
  {address} {Berlin Heidelberg New York},\ \bibinfo {year} {2001})\ pp.\
  \bibinfo {pages} {235--257}\BibitemShut {NoStop}%
\bibitem [{\citenamefont {Schwarz}(1985)}]{Schwarz:1985}%
  \BibitemOpen
  \bibfield  {author} {\bibinfo {author} {\bibfnamefont {K.~W.}\ \bibnamefont
  {Schwarz}},\ }\href {\doibase 10.1103/PhysRevB.31.5782} {\bibfield  {journal}
  {\bibinfo  {journal} {Phys. Rev. B}\ }\textbf {\bibinfo {volume} {31}},\
  \bibinfo {pages} {5782} (\bibinfo {year} {1985})}\BibitemShut {NoStop}%
\bibitem [{\citenamefont {Mason}\ \emph {et~al.}(2006)\citenamefont {Mason},
  \citenamefont {Berloff},\ and\ \citenamefont {Fetter}}]{Mason:2006}%
  \BibitemOpen
  \bibfield  {author} {\bibinfo {author} {\bibfnamefont {P.}~\bibnamefont
  {Mason}}, \bibinfo {author} {\bibfnamefont {N.~G.}\ \bibnamefont {Berloff}},
  \ and\ \bibinfo {author} {\bibfnamefont {A.~L.}\ \bibnamefont {Fetter}},\
  }\href {\doibase 10.1103/PhysRevA.74.043611} {\bibfield  {journal} {\bibinfo
  {journal} {Phys. Rev. A}\ }\textbf {\bibinfo {volume} {74}},\ \bibinfo
  {pages} {043611} (\bibinfo {year} {2006})},\ \Eprint
  {http://arxiv.org/abs/cond-mat/0605648} {arXiv:cond-mat/0605648} \BibitemShut
  {NoStop}%
\end{thebibliography}
%

\end{document}